# Gelation of Plasmonic Metal Oxide Nanocrystals by Polymer-Induced Depletion-Attractions

Camila A. Saez Cabezas[a], Gary K. Ong[a,b], Ryan B. Jadrich[a], Beth A. Lindquist[a], Ankit Agrawal[a], Thomas M. Truskett[a, c, 1], and Delia J. Milliron[a, 1]

[a] McKetta Department of Chemical Engineering, The University of Texas at Austin, Austin, Texas 78712, United States

[b] Department of Materials Science and Engineering, University of California, Berkeley, Berkeley, California 94720, United States

[c] Department of Physics, The University of Texas at Austin, Austin, Texas 78712, United States

[1] To whom correspondence should be addressed. Email: truskett@che.utexas.edu, milliron@che.utexas.edu

**Abstract**

Gelation of colloidal nanocrystals (NCs) emerged as a strategy to preserve inherent nanoscale properties in multiscale architectures. Yet available gelation methods still struggle to reliably control nanoscale optical phenomena such as photoluminescence and localized surface plasmon resonance (LSPR) across NC systems due to processing variability. Here, we report on an alternative gelation method based on physical inter-NC interactions: short-range depletion-attractions balanced by long-range electrostatic repulsions. The latter are established by removing the native organic ligands that passivate tin-doped indium oxide (ITO) NCs while the former are introduced by mixing with small polyethylene glycol (PEG) chains. As we incorporate increasing concentrations of PEG, we observe a reentrant phase behavior featuring two favorable gelation windows; the first arises from bridging effects while the second is attributed to depletion-attractions according to phase behavior predicted by our unified theoretical model. The NCs remain discrete within the gel network, based on X-ray scattering and high-resolution transmission electron microscopy. The infrared optical response of the gel is reflective of both the NC building blocks and the network architecture, being characteristic of ITO NC LSPR with coupling interactions between neighboring NCs.

**Introduction**

Nanocrystals (NCs), owing to their unique and highly tunable optical properties[1-5], hold promise as key constituents in next-generation optoelectronic materials and devices[1, 6-10]. Rich opportunities to enhance and diversify materials functionality motivate the development of multiscale NC architectures via bottom-up approaches[11] because the collective properties of NCs in assemblies depend on their organization. Nanoscale optical phenomena such as photoluminescence (PL) and localized surface plasmon resonance (LSPR) are especially responsive to electronic and electromagnetic coupling arising between NCs in close proximity. This effect is reflected in the optical properties of extended and dense NC assemblies with a high degree of inter-NC connectivity (*e.g.*, superlattices[12, 13] and films[14]), which deviate from those of their isolated components. Low density NC gel networks, where inter-NC bonding is constrained to moderate and controlled coupling, also have the potential to exhibit properties both dependent on their self-assembled architecture and reflective of their nano-sized building blocks. This potential has been realized for semiconductor quantum dot gels and aerogels[15], which exhibit excitonic PL red-shifted from the luminescence of isolated quantum dots due to energy migration through the gel network. However, nanoparticles (NPs) of plasmonic metals such as gold or silver have fused into wire-like networks when assembled into gels, obliterating the LSPR optical response characteristic of the isolated NPs[16, 17]. We sought a new strategy for gelation using physical bonding interactions to target gel assemblies of *discrete* LSPR-active metal oxide NCs, which we hypothesized could be maintained as discrete LSPR-active building blocks if assembled in this way. Our approach is not specific to the chemistry of the NCs employed and could potentially enable a broad class of gels assembled from diverse nanoscale components capable of reflecting their individual properties.

Briefly, gelation is achieved by balancing attractions and repulsions in colloidally stable NC dispersions. In previously published examples, these interactions are simultaneously tuned by progressive oxidative ligand removal or controlled chemical bridging between surface bound species and linking agents (*e.g.* ions or molecules). The former has been adapted across noble metal[16, 18-21], metal chalcogenide[22-26], and metal oxide[27-31] systems, but this method is prone to fuse NCs during assembly and consequently limits the realization of size and shape-dependent NC optical properties (*i.e.*, PL and LSPR) within the gels. While gelation via chemical bridging is a viable strategy to mitigate NC fusing, translating this approach across NC materials requires customizing surface functional groups for specific NC compositions, so far limited to metal chalcogenide NCs[32, 33] and Au NPs[17]. However, even this approach did not prevent fusing of Au NPs with a concomitant loss of LSPR response. Among the sparse reports on metal oxide NC gels, gelation has been most often achieved by fusion upon ligand-removal[29-31] or by triggering the entanglement of concentrated anisotropic NPs: chains of pre-destabilized titania ($TiO_2$) NC[28, 34-36] form gels upon heating while tungsten oxide ($W_{18}O_{49}$) nanowires[37] and yttria ($Y_2O_3$) nanosheets[38] form gels upon centrifugation. Once more, this approach offers limited control over gel structure and thereby the associated properties, and it cannot be easily generalized to assemble discrete isotropic metal oxide NCs.

In light of these limitations, we were motivated to develop an alternative route for NC gelation based upon non-specific physical interactions by combining depletion-attractions and electrostatic repulsions. Previous studies have demonstrated that this combination can help drive the gelation of polymer colloids[39-41] and the assembly of proteins (into gels[42], clusters[43], and crystals[44, 45]), and hence it holds potential for tunable gelation of NCs. Conceptual understanding of the strength and the range of depletion attractions requires consideration of only a few parameters: the concentrations of the primary colloid and the depletant (smaller, weakly interacting co-solute) and their relative dimensions. The addition of long-range repulsions to depletion interactions can favor "open" gel structures as opposed to dense colloidal phases[46-48]. Therefore, a method to controllably introduce repulsive forces, here electrostatics, is needed to realize NC gelation via physical depletion-attractions.

In this study, we demonstrate the polyethylene glycol (PEG)-mediated gelation of tin-doped indium oxide (ITO) NCs stripped of their native ligand shell. Electrostatic stabilization enables inter-NC repulsions while introducing short-chain PEG triggers depletion-attractions. We investigate the influence of PEG concentration ([PEG]) on competing inter-NC interactions for a fixed ITO volume fraction, and we observe two gelation thresholds at distinct [PEG], each preceded by a fluid regime (*i.e.*, flowing dispersion) of discrete ITO NC clusters. Since PEG is known to adsorb onto acidic metal oxide surfaces, we attribute the emergence of a first gelation window at low [PEG] to bridging of neighboring NCs by PEG chains while the higher [PEG] gelation window is attributed to depletion-attractions. To support our assertion and assess the gelation mechanism, we compare our experimental results to thermodynamic phase behavior predictions from a unified theory formulated to capture polymer-mediated bridging and depletion-attractions. Because depletion gelation effectively avoids the fusion of ITO NCs and uncontrolled aggregation, we achieve an optically active and transparent gel with LSPR similar to that of the isolated ITO NCs, but shifted and broadened by inter-NC coupling. To explore influence of assembly on optical properties, we perform far-field and near-field electromagnetic simulations based on structural information extracted from small-angle X-ray scattering (SAXS). Our simulation results predict near-field enhancement manifested as "hot spots" within the gel network that may be leveraged in future studies for energetic coupling between LSPR and molecular vibrational modes[49], or other optical transitions[50].

**Results and Discussion**

ITO NCs with an average radius $R_{ITO}$ of 2.83 ± 0.36 nm (SI, Fig. S1 illustrates sizing analysis based on scanning transmission electron microscopy (STEM)) were synthesized following a procedure adapted from the literature[51, 52] (see Materials and Methods). Charge-stabilized ITO NCs are obtained by chemically removing the native oleate/oleylamine ligands on the NC surface in the presence of nitrosonium tetrafluoroborate[53] (see Materials and Methods). During this process, the NCs undergo a phase transfer from hexane to a polar aprotic dimethylformamide (DMF) phase without any signs of etching or morphology change (Fig. 1a and b). The integrity of the ITO NC cores is confirmed with dynamic light scattering (DLS) since the measured hydrodynamic diameter of ITO in DMF is 6.5 nm, close to the diameter of the NCs measured by STEM (SI, Fig. S2). Ligand-stripping from the ITO surface is assessed with Fourier transform infrared (FTIR) spectroscopy and zeta potential measurements. As shown on Fig. 1c, the characteristic oleic acid and oleylamine bands corresponding to $-CH_2$ (2922 and 2851 cm$^{-1}$) and $=CH$ (3005 cm$^{-1}$) vibrational modes[54] are absent in the FTIR spectrum of ligand-stripped ITO. Unlike ligand-capped ITO NCs dispersed in non-polar solvents, the bare NC dispersion exhibits strongly positive zeta potentials in both acetonitrile (ACN) and DMF while retaining colloidal stability (Fig. 1d and SI, Fig. S2). This change in NC surface charge has been previously attributed to uncoordinated metal cations exposed upon removal of anionic ligands[53, 55].

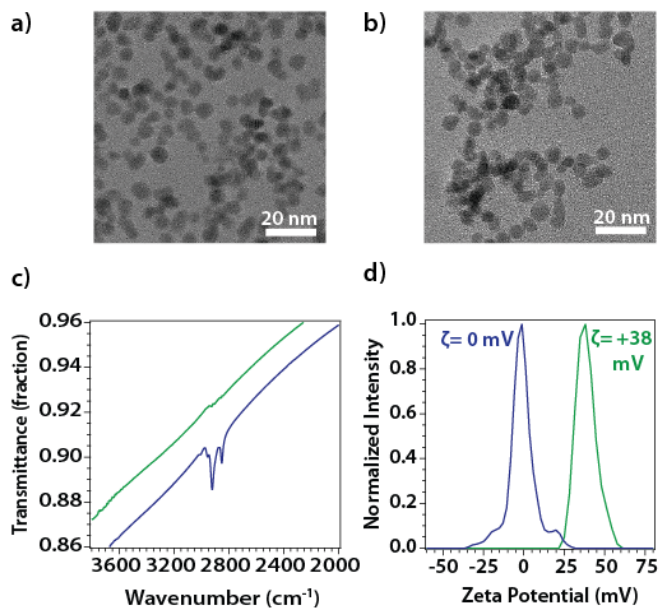

**Figure 1. Surface modification of ITO NCs.** a) Transmission electron microscopy (TEM) of as-synthesized ITO NCs, b) TEM of ligand-stripped ITO NCs, c) FTIR comparison of as-synthesized (blue) and ligand-stripped (green) ITO NCs. The sloping profile in the spectra corresponds to absorption due to LSPR, and d) zeta potential comparison of as-synthesized ITO NCs dispersed in hexane (blue) and ligand-stripped ITO NCs dispersed in ACN (green).

Our NC gelation strategy leverages physical bonds formed by balancing long-range electrostatic repulsions due to surface charge and short-range attractions induced by depletants, aiming to create stable open, arrested, and percolated networks. Short PEG polymer chains ($M_n$= 1100 g/mol) were selected as depletants based on the following criteria: the need for a co-solute with a radius of gyration, $R_{g\,PEG}$, smaller than $R_{ITO}$, PEG's ability to raise the osmotic pressure in solutions[56], and PEG's compatibility with polar aprotic solvents. Although PEG dissolves in both DMF and ACN, the latter has a Hildebrandt solubility parameter closer to the one of PEG[57, 58] and was therefore chosen as the solvent matrix for our ITO-PEG gels. In fact, PEG precipitates in DMF within a couple of days at concentrations of 664 mg/ml while no signs of precipitation are observed for PEG in ACN solutions of the same concentrations for over one year (SI, Fig. S3). Moreover, we estimated $R_{g\,PEG}$ of these PEG chains in ACN to be 0.98 nm from SAXS sizing analysis (SI, Fig. S4) to ensure that the depletant size criterion would be fulfilled. This value is in good agreement with the expected $R_{g\,PEG}$ of PEG ($M_n$= 1100 g/mol) from literature[59]. Previous studies on polymer-induced depletion-attractions have shown that the strength of the attraction is tunably increased as a function of depletant concentration, which in turn dictates the extent of the network's connectivity[40, 46], affecting gel structure and any properties dependent on the local environment and valence of the NCs in the network. Since colloidal NC depletion gels have not been previously reported, the conditions to induce gelation were discovered by varying the amount of PEG for a constant ITO volume fraction. Here, we introduce various amounts of PEG into charge-stabilized ITO NC dispersions in ACN of fixed volume fraction (see Materials and Methods). Experimentally, as we progressively increase the PEG concentration at fixed ITO NC volume fraction (3.84 vol. % from inductively coupled plasma-atomic emission spectroscopy (ICP-AES), SI, Table S1), we observe a fluid (*i.e.*, flowing dispersion) up to a first threshold for gelation at [PEG]= 46.0 mM, then reentrant behavior back to a flowing dispersion, followed by a second occurrence of a gel at [PEG]= 534 mM (Fig. 2 insets).

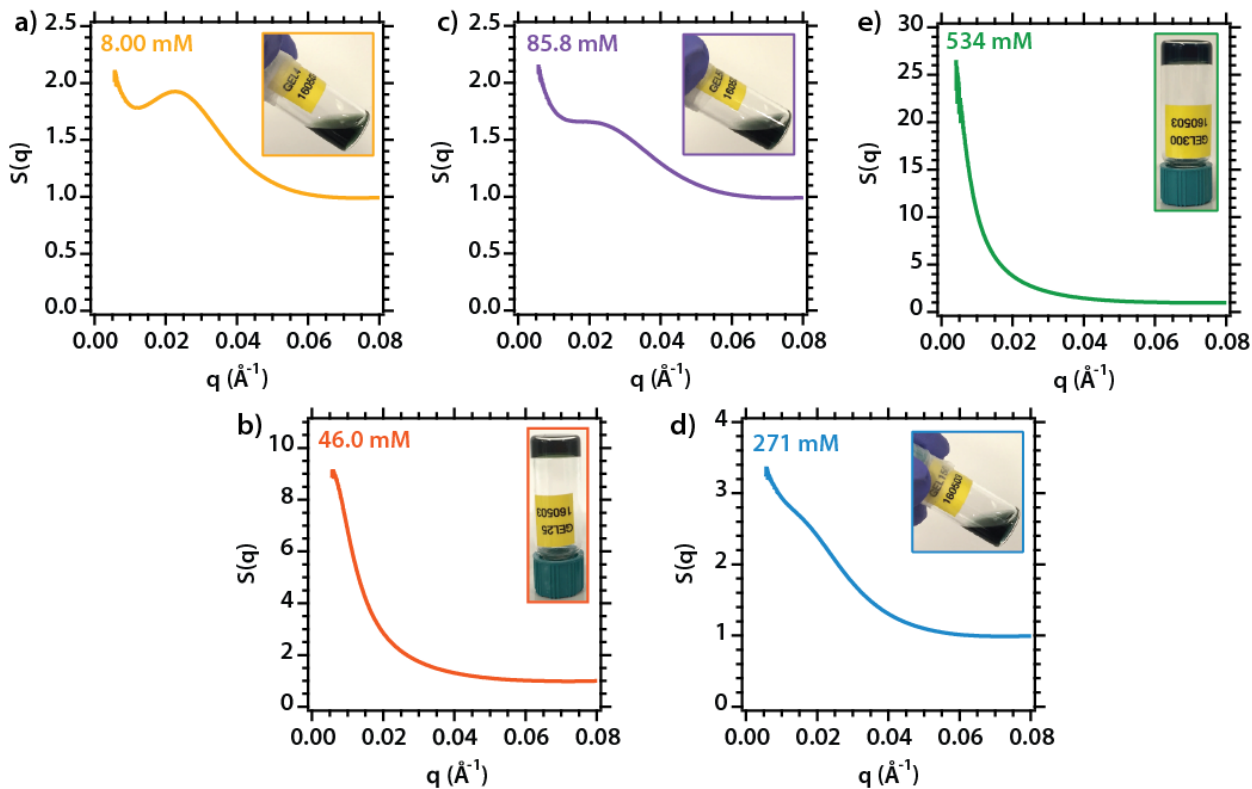

**Figure 2. SAXS characterization of the ITO-PEG flowing dispersions and gels.** a) Structure factor S(q) of flowing dispersion with [PEG]= 8.00 mM, b) S(q) of gel with [PEG]= 46.0 mM, c) S(q) of flowing dispersion with [PEG]= 85.8 mM, d) S(q) of gel with [PEG]= 271 mM, and e) S(q) of flowing dispersion with [PEG]= 534 mM. S(q) plots are accompanied with photographs insets of the corresponding ITO-PEG mixture.

To investigate the conditions that enabled gelation and to characterize the fluid regime, the ITO-PEG mixtures formed at different [PEG] were probed with SAXS (see Materials and Methods).

Specifically, we examine the structure factor S(q) as a function of [PEG] by removing the form factor contribution to the SAXS data (see SI for more details and Fig. S5) to gain insight into the physical origins of the self-assembly. As a result, we identify two distinct interaction regimes (Fig. 2). First, we note a prominent S(q) peak (q~0.023 Å$^{-1}$, Fig. 2a) for the ITO-PEG flowing dispersion of lowest [PEG] emerging at a lower q than that indicative of NC-NC correlation (see SI, Fig. S6 for full q range S(q)), thus characteristic of intermediate range order (IRO) in colloidal assemblies[48, 60, 61]. Considering the IRO behavior of this ITO-PEG mixture ([PEG]=8.00 mM), we establish that competing short-range attractions and long-range repulsions frustrate large scale aggregation and lead to the formation of discrete ITO NC clusters (~270 Å length scale) dispersed in ACN. In addition, we observe similar IRO behavior for the [PEG]= 85.8 and 271 mM flowing dispersions. Although the S(q) peaks near q~0.02 Å$^{-1}$ for the latter dispersions are less pronounced and broadened (Fig. 2c and d), likely due to an increase in attraction strength and cluster polydispersity, the presence of dispersed discrete ITO clusters is still apparent. Second, in all cases, we observe that S(q) diverges as q approaches zero suggesting systems dominated by attractions and thermodynamic compressibility[48, 60]. In particular, the S(q) intensity at the lowest resolvable q is approximately an order of magnitude higher when gelation occurs compared to the S(q) intensity of all flowing cluster dispersions (Fig. 2b and e). Prior colloidal assembly studies[47, 62-64] have reported a comparable S(q) intensity increase (of an order of magnitude or larger) when a colloidal system transitions from a fluid state to a gel through spinodal decomposition.

Further inspection of S(q) for both low and high [PEG] gels by employing Beaucage's unified function[65-67] approach for complex structures provides insight into the structural hierarchy of the gels and their respective fractal dimension ($D_f$). First, the number of apparent structural length scales along with their respective limits are determined from a derivative analysis (see SI, Fig. S7). Following this method, we identify two distinct structural length scales in the S(q) of the high [PEG] gel whereas the S(q) of the low [PEG] gel only exhibits one. Thereafter, our derivative analysis results guide the S(q) unified fitting of both gels (see SI, Fig. S8). While the S(q) of the [PEG] = 46.0 mM gel scatters as a mass fractal of $D_f$ = 2.2, the S(q) of the [PEG]= 534 mM gel is composed of a mid-q (0.04 Å$^{-1}$ < q < 0.058 Å$^{-1}$) scattering contribution from ITO clusters plus a low-q (0.01 Å$^{-1}$ < q < 0.02 Å$^{-1}$) scattering contribution attributed to the presence of a percolated ITO gel network ($D_f$ = 2.09). For both low and high [PEG], we associate gelation with slow bonding kinetics since the fitted $D_f$ values fall within the expected range (2.0-2.2) for reaction-limited cluster aggregation (RLCA) systems[68-70].

In the SAXS spectra of all [PEG], we observe the form factor of the discrete NCs, with fitting results yielding constituent particles with a radius of 2.95, 2.89, and 2.82 nm in the case of the lowest [PEG] dispersion, low [PEG] gel, and high [PEG] gel, respectively. This consistency suggests that the NCs remain discrete under all PEG-induced assembly conditions. The persistence of the discrete NCs is further supported by high-resolution transmission electron microscopy (HR-TEM) imaging of a dried gel with [PEG] = 534 mM (SI, Fig. S9), where individual NCs are discerned, without apparent crystallographic continuity between adjacent NCs that would indicate fusing of NCs in the network. Scherrer analysis of X-ray diffraction (XRD) complements our observations by HR-TEM, where the crystallite size of the ligand-stripped NCs (6.19 nm for the (222) peak) is found to be comparable to the ITO crystallite size in the [PEG]= 534 mM gel (5.89 nm for the (222) peak, see SI, Fig. S10), both of which are in turn consistent with diameters measured by STEM, SAXS, and DLS. Preventing NC fusion throughout the assembly process is particularly advantageous for depletion gelation since the attraction strength can be weakened by reducing the depletant concentration relative to the primary particle and, in principle, reverse gelation. We achieved the disassembly of the ITO-PEG gel ([PEG]= 534 mM) by adding 600 μL of ITO-PEG flowing dispersion ([PEG]= 8.00 mM) to dilute [PEG] in the mixture by a factor of three at a fixed ITO vol. % (Movie S1). A stable flowing dispersion is recovered by gentle manual agitation without using sonication or vortexing.

Considering the [PEG]-dependent phase progression including a reentrant regime, we hypothesized that ITO NC assembly might be influenced by inter-NC attractions other than depletion-attractions. Specifically, we propose that, in addition to its depletant role, PEG can bridge adjacent NCs by adsorbing onto ITO surfaces. Because PEG chains are known to preferentially adsorb on acidic oxide surfaces (the isoelectric point (IEP) of the ITO NCs used in this work is between 4 and 5, SI, Fig. S11) via hydrogen bonding and subsequently aggregate oxide particles[71-73], we deduce that low PEG concentrations (*e.g.,* [PEG]= 46.0 mM) can favor bridging gelation. It is worth noting that PEG adsorption on the ITO surface does not hinder the inter-NC long-range electrostatic repulsion necessary to form open depletion gels since ITO-PEG dispersions in ACN still exhibit a strong positive zeta potential (SI, Fig. S12). In this light, prior work by Luo, Zhao, and co-workers[40, 74] described an analogous experimental phase progression in a polystyrene microsphere system where bridging and depletant effects are both operative. They showed the emergence of the following phase transition sequence as the concentration of the smaller adsorbing species (poly(N-isopropylacrylamide) or PNIPAM) in the system increases: bridging-induced aggregation → stabilization of discrete microspheres → depletion-induced aggregation. Moreover, they determined that since depletion-attraction interactions are only favored once the adsorbing molecules have saturated the colloidal surface and bridging attractions are hindered, the assembly mechanism (*i.e.*, bridging or depletion) is highly sensitive to changes in the colloid-to-adsorbing molecule concentration ratios.

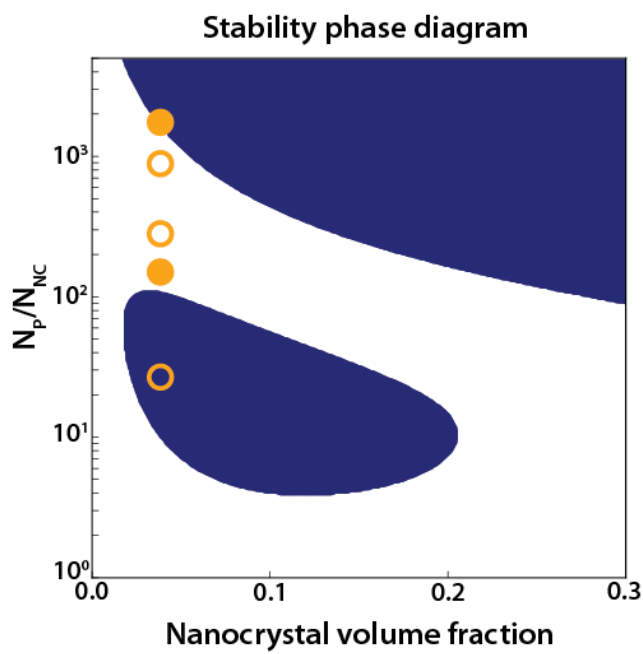

**Figure 3. Comparison of theoretical model to experimental observations.** Theoretical phase diagram overlaid with experimental data points. Open circles represent experimental flowing dispersions and closed circles represent experimental observation of gels. Regions where bridging and depletion gelation are predicted to occur are delimited by lower and upper blue areas, respectively.

To assess our proposed mechanism for reentrant gelation in ITO-PEG dispersions, we devised a theoretical model that is unique in possessing a unified description of bridging and depletion effects. The free-energy theory synergistically combines a well-accepted theoretical treatment for the Asakura Oosawa (AO) model (depletion) with the accurate Wertheim theory for strong association (bridging). Various physical parameters enter the theory: the NC to polymer diameter ratio ($d_{NC}/d_P$), the number of polymers that can adsorb onto the NC surface before saturation ($n_{ads}$), the number of NCs that a single polymer chain can bridge ($n_{bind}$), and the polymer-NC thermal adsorption volume (v), which encapsulates the combined effects of adsorption energy, temperature and the spatial range of the attraction (see SI for more details). To specifically model our NC gels, for $d_{NC}/d_P$ we use the experimental value of ~ 3, and for $n_{bind}$ we use the physically reasonable value of 2 based on the short PEG chains employed and ITO. For v and $n_{ads}$ we explored various possibilities and the associated phase behavior, one example of which is shown in Fig. 3 for $n_{ads}$=30 and v= 0.181 yielding a reentrant (liquid→gel→liquid→gel) phase diagram that is almost quantitatively in accord with the experimental results. Given our choices above for $n_{ads}$ and $n_{bind}$, zero-temperature mean-field theoretical calculations[75] indicate that the bridging regime should not exceed a polymer-to-colloid ratio of 435 at any volume fraction or value of v. Therefore, the first spinodally unstable regime with increasing depletant is driven by NC-polymer bridging (which saturates upon surface coating) and the second by depletion. Importantly, the phase diagram is always qualitatively the same for physically reasonable values of $n_{bind}$: bridging gels form when the ratio of the number of polymers per NC ($N_P/N_{NC}$) is of order 10-100 and whereas of order 1000 is required for depletion—as seen in the experiments.

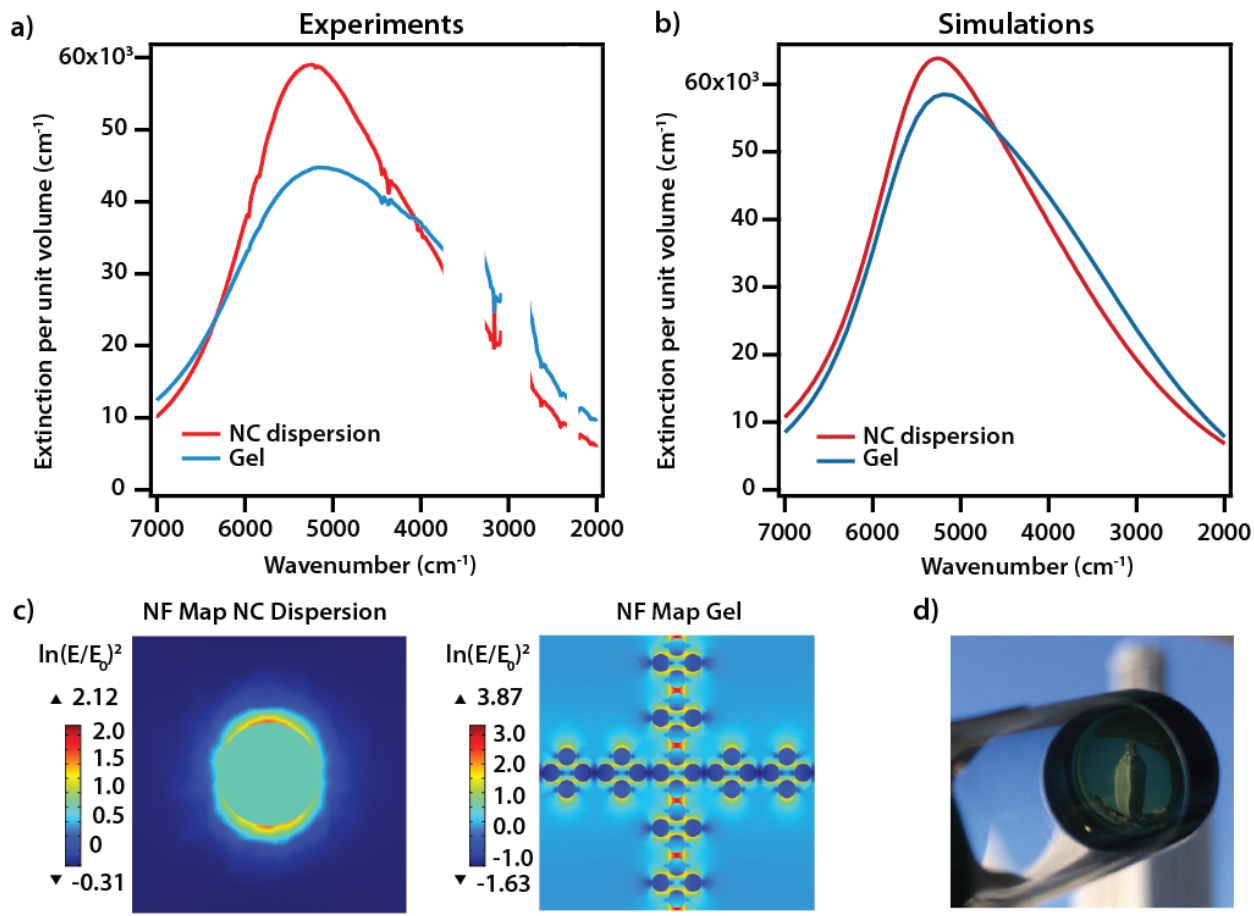

**Figure 4. Experimental and simulated optical properties of ITO-PEG depletion gel.** a) Experimental extinction spectra of ITO depletion gel ([PEG]=534 mM, blue) and ligand-stripped ITO NC dispersion (red), b) simulated extinction spectra of ITO depletion gel and ligand-stripped ITO NC dispersion, c) simulated near-field (NF) maps of ITO depletion gel and ligand-stripped ITO NC dispersion, and d) photograph of highly transparent ITO depletion gel under natural lighting against University of Texas Tower.

Our depletion-attraction assembled ITO-PEG gel formed at the highest [PEG] is optically active and exhibits an extinction spectrum reminiscent of that of the discrete ITO NC building blocks. As shown in Fig. 4a, the LSPR peak of the gel is slightly red-shifted (by 101 cm$^{-1}$) from that of dispersed ITO NCs in ACN. We attribute the similarity between these spectra to our successful preservation of the NC's integrity as they are integrated in the gel network, avoiding inter-NC fusion (see SI, Fig. S9-S10). The modest red-shift of the gel LSPR, its reduced peak intensity, and significant broadening towards lower energies compared to the spectrum of isolated NCs are all characteristic of LSPR coupling between NCs, as previously studied in extended assemblies, such as films, of colloidal metal NPs[76-78] and metal oxide NCs[49, 79, 80]. Therefore, the perturbations in the gel LSPR suggest structure-dependent coupling interactions between ITO NC nearest neighbors.

To confirm the influence of structure on the gel's far-field optical properties and to anticipate their near-field optical properties, we simulated the optical response of an extended network composed of PEG-coated ITO octahedral clusters (10 nm in radius) with a NC volume fraction of 3.88 vol. %, similar to that we studied experimentally (see SI for more details). Although an idealized structural representation was used to ensure computational tractability, the simulated network was designed to approximate the experimentally measured NC volume fraction in the gel and the structural hierarchy ascertained from analysis of the SAXS data described earlier. As shown in Fig. 4b, the differences in the simulated extinction spectrum of the gel configuration and the ITO NC dispersion are in qualitative agreement with the ones found experimentally (Fig. 4a) and thus reflect NC-NC coupling interactions facilitated by the assembly of an extended gel network. Based on the simulations, we anticipated that such coupling effects should give rise to intense "hot spots" of greatly enhanced electric field intensity confined between NCs in the gel assemblies where the electromagnetic near-fields of neighboring NCs overlap. Simulated near-field maps shown in Fig. 4c demonstrate the near-field enhancement under resonant optical excitation of the gel network. Altogether, these findings highlight that depletion-attraction gelation of discrete NCs leverages the LSPR sensitivity to the local NC arrangement to extend the NC's intrinsic IR plasmonic range and diversify the optical response without suppressing the LSPR, and it allows generation of localized "hot spots", thereby providing coupling opportunities to other optical transitions relevant to surface enhanced infrared absorption spectroscopy (SEIRA) and sensing applications[10, 49, 81] otherwise inaccessible in networks of fused NCs.

**Conclusion and Outlook**

The strategy described here demonstrates the potential for tunable gels based on reversible physical bonds, and with responsive optical properties. A basic requirement for optical materials is that scattering does not interfere with the absorption, reflection, and luminescence properties of interest. Our depletion-attraction strategy produces a highly transparent ITO gel that remains stable for over one year without developing haze noticeable to the eye (Fig. 4d). Obtaining transparent self-supported NC gels has remained a challenge in the field since most established gelation methods give rise to fast-growing networks of large aggregates (scatterers) that ultimately form opaque gels. Instead, as alluded to earlier for our system, competing electrostatic repulsions and depletion-attractions favor slow bonding kinetics (RLCA) and thereby facilitate the formation of fractal aggregates with characteristic length scales smaller than the wavelength of visible light ($R_g$ = 53 nm, SI, Fig. S8). Our results are consistent with a previous study by Korala and Brock[82] on the aggregation kinetics of CdSe NCs in which it was determined that simultaneously accessing the RLCA regime and suppressing large scale aggregation are necessary conditions to induce transparency in a NC gel. Accordingly, gaining insight into the interplay between inter-particle interactions, structure, and aggregation kinetics is key to rationalize and exploit the NC gel properties stemming from nanoscale building blocks.

More generally, we showed how the combination of depletion-attractions and electrostatic repulsions can realize the assembly of ITO NC gels. While the addition of PEG mediates inter-NC attractions, competing long-range electrostatic repulsions resulting from ligand stripping encourage the formation of a self-supported gel rather than a dense and collapsed material. This model system was designed to investigate depletion-attraction interactions in colloidal NCs and develop a gelation strategy that should in principle be broadly applicable across NC systems. We observed the emergence of two gelation windows interspaced with flowing dispersion states of discrete ITO clusters. Transitioning from a flowing to a solid-like gel state was accompanied by a strong S(q) divergence and intensity increase at the lowest resolvable q, a characteristic of colloidal aggregation through spinodal decomposition reported in prior literature. In this regard, our theoretical phase behavior predictions, based on a unified bridging and depletion-attraction description that captured PEG's affinity for oxide surfaces and ability to bridge adjacent ITO NCs, supported our experimental results and identified two spinodally unstable regions favoring bridging and depletion gelation at low and high [PEG], respectively. Moreover, we structurally differentiated the dominant assembly mechanism in each gel since we recognized two scattering length scales (NC < fractal gel) in the bridging gel as opposed to the three scattering length scales (NC < cluster < fractal gel) apparent in the depletion gel likely due to depletion-attractions acting on pre-assembled ITO clusters instead of discrete NCs.

From an application perspective, we demonstrated the effectiveness of our depletion-attraction approach to produce an optically active gel featuring LSPR representative of the ITO NC building blocks by retaining the discrete NC morphology in the network, which has not been achieved with previously reported gel processing methods for plasmonic NCs. By controlling the extent of aggregation in a RLCA regime, our gelation approach favored the formation of a transparent gel, showing no signs of visible haze or scattering. In addition, we compared our optical spectra from experiments to electromagnetic simulations to highlight near-field enhancement opportunities generated by nearest-neighbor coupling effects in the gel network, a promising feature that encourages further exploring NC gels as an alternative material for coupling applications. Finally, we envision that, extending our gelation approach to other NC systems could motivate further studies to improve our insight on structure-property relationships in assemblies to thus achieve systematic design of NC gel properties. We believe that our framework could also contribute to the development and diversification of multicomponent NC gels as a means to unlock even more complex NC gel functionality.

**Materials and Methods**

*Materials.* All chemicals were used as received and without further purification. Indium acetate (In(ac)$_3$, 99.99% trace metal basis), tin (II) acetate (Sn(ac)$_2$), oleylamine (technical grade, 70%), oleic acid (technical grade, 90%), N,N-dimethylformamide (DMF, ACS reagent, $\geq$ 99.8%), nitrosonium tetraborofluorate (NOBF$_4$, 95%), acetonitrile (ACN, ACS reagent, $\geq$ 99.5%), nitric acid (ACS reagent, 70%), and hydrochloric acid were purchased from Sigma Aldrich. Hexane (ACS reagent, various methylpentanes 4.2%, $\geq$ 98.5%), reagent alcohol (ethanol 88-91%, methanol 4.0-5.0%, Isopropyl Alcohol 4.5-5.5%), and toluene ($\geq$ 99.5%) were purchased from Fisher Scientific. Polyethylene glycol (PEG, Mn=1100 g/mol) was purchased from Polymer Source.

*ITO Nanocrystal Synthesis and Ligand-Stripping.* ITO NCs were synthesized in an air-free environment using a standard Schlenk line technique following a procedure adapted from previous literature reports[51, 52]. Briefly, 2.5 g (8.6 mmol) of In(ac)$_3$ and 0.225 g (0.95 mmol) of Sn(ac)$_2$ were mixed with 10 mL of oleylamine, degassed under vacuum for 1 hour at 120 °C, and then heated to 230 °C under inert nitrogen atmosphere for 1 hour to nucleate and grow the NCs. After the reaction, the mixture was diluted with 5 mL of hexane and 1 mL of oleic acid and transferred to a centrifuge tube. The resulting ITO NCs were recovered and purified by performing five cycles of precipitation with reagent alcohol, centrifugation, and re-dispersion in hexane.

To remove the hydrophobic ligands bound to the ITO surface, 60 mg of $NOBF_4$ were added to a two-phase mixture containing equal volumes of DMF (2 mL) and ITO dispersion in hexane (50 mg/ml, 2 mL) following a procedure reported previously[53]. The mixture was sonicated for 1 hour to promote the transfer of bare ITO NCs to the DMF layer. After a successful ligand-stripping procedure, the hexane layer turns clear and the DMF layer has a blue/green color. After discarding the hexane layer, the ligand-stripped ITO were purified by performing seven cycles of precipitation with toluene, centrifugation, and re-dispersion in DMF.

***Ligand-Stripping Procedure Characterization.*** NC hydrodynamic diameter and zeta potential before and after ligand-stripping were measured with a Malvern Zetasizer Nano ZS. Samples were prepared by diluting the NC dispersions to ~1 mg/ml and filtering them through a PTFE membrane. DLS samples were placed in a disposable plastic micro cuvette (ZEN0040, Malvern) while zeta potential samples were placed in a glass cuvette and measured with a dip cell (ZEN1002, Malvern). The IEP of ligand-stripped ITO NCs was determined by measuring zeta potential as a function of pH using Malvern's MPT-2 autotitrator. Ligand-stripped ITO NCs were dispersed in 1 mM hydrochloric acid aqueous solution and titrated with a 0.1 M sodium hydroxide solution followed by a titration with a 0.1 M hydrochloric acid solution in a folded capillary zeta cell (DTS1070, Malvern). FTIR spectra of ITO NCs before and after ligands-stripping were recorded with a Burker-Vertex 70/70v spectrometer. The films were dropcasted from dilute dispersions (~ 1 mg/ml) on calcium fluoride IR-transparent windows.

***ITO-PEG Assemblies***. Purified ligand-stripped ITO NCs were dispersed in 8.00 mM (based on $M_n$=1100 g/mol) PEG in ACN solution (final concentration 3.84 % vol. by ICP-AES). The solution was stirred at 600 rpm for 48 hours to ensure colloidal stability and homogeneity. This mixture was used as a stock dispersion to obtain the higher [PEG] assemblies. Varying amounts of PEG were added to 300 μL aliquots of the ITO-PEG stock dispersion to form flowing dispersions or gels: 12.48 mg (final [PEG]= 46.0 mM), 25.62 mg (final [PEG]= 85.8 mM), 86.69 mg (final [PEG]= 271 mM), 173.43 mg (final [PEG]= 534 mM). After the final amount of PEG was added, each mixture was sonicated for one minute to dissolve the PEG. The dispersions were kept in sealed vials and remained unperturbed during the self-assembly process.

***Microscopy (TEM and STEM).*** Samples were prepared by dropcasting 5 μL of dilute NC dispersion onto a copper grid (Pelco® ultrathin carbon-A 400 mesh, Ted Pella). The ITO-PEG gel ([PEG]= 534 mM) was freeze-dried (immersion in liquid nitrogen for 1 min followed by vacuum drying for 15 min) and diluted with ethanol before dropcasting onto a copper grid. TEM images were captured on a JEOL 2010F TEM with a Schottky Field Emission source operated at 200 kV. Low resolution images used for NC sizing analysis were captured using a Hitachi S5500 in STEM mode at a 30 kV accelerating voltage.

***ICP-AES Measurement***. The overall tin dopant concentration of the as-synthesized ITO NCs and the volume fractions of ITO dispersed in pure ACN (used for optical properties analysis in Fig. 4) and ITO gel ([PEG]=534 mM) were determined from ICP-AES data collected with a Varian 720-ES ICP Optical Emission Spectrometer. The samples were digested in 70 wt. % nitric acid for 36 hours. Measured standards and samples were diluted to 2 vol. % nitric acid in ultrapure Milli-Q water.

***SAXS Characterization***. SAXS measurements were performed at the Lawrence Berkeley National Laboratory Advance Light Source (ALS) beamline 7.3.3 at 3.8 m sample-detector distance. A silver behenate standard[83] was used to calibrate the scattering spectra. All ITO-PEG samples were enclosed in flame sealed glass capillaries (Charles-Supper Company, Boron Rich, 1.5 mm diameter, 0.01 mm wall thickness) and measured in transmission configuration. Capillaries containing neat ACN were used for background subtraction. The Igor Pro-based Nika software[84] for two-dimensional (2D) data reduction was used for instrument calibration and to convert 2D detector data into 1D data by circular averaging. Additional data processing and analysis details are provided in the SI.

***XRD Characterization***. XRD patterns were collected on a Rigaku R-Axis Spider using Cu K$_\alpha$ radiation. The ligand-stripped ITO and ITO-PEG gel ([PEG]= 534 mM, identical sample measured by SAXS) were enclosed in glass capillaries and were measured in transmission configuration. Crystallite sizes were determined by Scherrer analysis and a lanthanum hexaboride (LaB$_6$) reference provided by the National Institute of Standards and Technology (NIST) was used to correct for instrumental broadening (see SI, Fig. S10).

***Unified Theoretical Bridging and Depletion Gelation Model.*** Phase diagram predictions utilize an approximate free energy expression for the Asakura Oosawa (AO) model modified to include short-range polymer-NC surface attractions. Formally, the free energy is decomposed as $a \equiv a_{AO} + a_B^{(ex)}$ where $a_{AO}$, and $a_B^{(ex)}$ are the AO and excess bonding (from polymer-NC adsorption) free energy contributions. For $a_{AO}$ we employ the theory of Lekkerkerker et al.[85] and for $a_B^{(ex)}$ we use Werthiem first order association theory[86, 87]. Spinodal boundaries are identified from the Hessian matrix ($H$) of partial derivatives with respect to polymer and NC densities by the satisfaction of $\det H \leq 0$. For more details, please see SI Section S2.

***LSPR Measurement and Analysis.*** Absorption spectra were collected on an Agilent-Cary 5000 UV-Vis-NIR spectrophotometer and a Burker-Vertex 70/70v FTIR spectrometer. ITO NC dispersions at various volume fractions in pure ACN and the ITO-PEG gel were measured in a IR transparent liquid cell with $CaF_2$ windows of 0.5 and 0.02 mm path length, respectively. To calculate the extinction cross section, the Beer-Lamberts law was used. To simulate the optical property of the ITO-PEG gel ([PEG]= 534 mM), an idealized PEG-coated ITO network was designed using the design module in COMSOL. For more details, please see SI Section S3.

**Acknowledgements**

This research was supported by the National Science Foundation (NSF) through the Center for Dynamics and Control of Materials: an NSF MRSEC under Cooperative Agreement DMR-1720595. SAXS data were collected at the Advanced Light Source's beamline 7.3.3 at the Lawrence Berkeley National Laboratory, a user facility supported by the U.S. Department of Energy (DOE) Office of Science under contract no. DE-AC02CH11231. G.K.O was supported in part by a NSF Graduate Research Fellowship under grant number DGE-1106400. Further support is acknowledged from the Welch Foundation (F-1848 and F-1696 and the NSF (CHE-1609656 and CBET-1247945). The authors would also like to acknowledge Angela M. Wagner for her assistance with IEP measurement and Corey M. Staller for his assistance with ICP measurements.

*Molecular Physics* 65(5):1057–1079.

**TOC Figure**

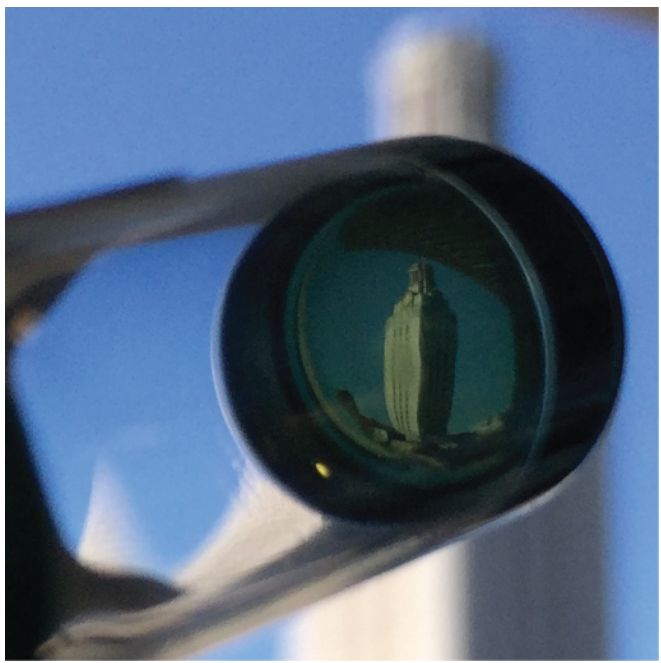